\documentstyle[11pt,newpasp,twoside,epsfig]{article}
\def\edcomment#1{\iffalse\marginpar{\raggedright\sl#1\/}\else\relax\fi}
\reversemarginpar

\begin{document}
\title{SZ surveys with the Arcminute MicroKelvin Imager}
 \author{Michael E. Jones}
\affil{Cavendish Laboratory, Madingley Road, Cambridge, CB3 0HE, United Kingdom}

\begin{abstract}
The Arcminute MicroKelvin Imager (AMI) is an instrument currently
under construction in Cambridge designed to produce a survey of galaxy
clusters via the Sunyaev-Zel'dovich effect. It consists of two
interferometric arrays, both operating at 12--18 GHz; one of ten 3.7-m
antennas to provide good temperature sensitivity to arcminute-scale
structures, and one of eight 13-m antennas (the present Ryle
Telescope) to provide flux sensitivity for removing contaminating
radio sources. The telescope is due to be observing by 2003, and will
produce a public survey of galaxy clusters, as well as being available
to guest observers.

\end{abstract}

\section{Introduction}

It is now widely recognised that an important next step in cosmology
is to conduct a survey of galaxy clusters via the SZ effect. Galaxy
clusters are the largest gravitationally bound objects in the
universe, and by pointing out the peaks of the initial density field,
they provide a sensitive indicator of the growth of structure and the
parameters that control it (e.g. Bartlett \& Silk 1994). They are also
large enough to provide a fair sample of the material consitituents of
the universe. In order to fully exploit the potential of clusters for
learning about the universe, however, it is necessary to be able to
select them according to uniform intrinsic properties over a large
range in redshift. The promise of SZ surveys is that they can in
principle select clusters over a very large redshift range with
well-understood and physically meaningful selection criteria (Barbosa
et al 1996, Holder et al 2000, Kneissl et al 2001).

The problem is that until now the observations required have been
exceedingly difficult. The SZ effect is faint, and existing telescopes
have only had sufficient sensitivity and sky coverage to study the
most massive clusters, whose existence was previously known from
optical or X-ray surveys. A new generation of instruments is now in
prospect that will provide a qualitative step in sensitivity and sky
coverage to finally fulfil the promise of the SZ technique. Here we
will describe one such instrument currently under construction, the
Arcminute MicroKelvin Imager, AMI.

\section{Interferometers for SZ observation}

Although the first detections of the SZ effect were made using
single-dish radiometers (e.g. Birkinshaw 1984), the majority of the
detections made so far have been using interferometers (eg Jones et al
1993, Carlstrom, Joy, \& Grego 1997). Interferometers offer several
significant advantages over total power measurements: they are not
susceptible to scan-synchronous systematics; they can use the rotation
of the earth to modulate the sky signal in a way that is
distinguishable from other signals such as crosstalk or groundspill;
they are insensitive to most atmospheric emission; and they can
simultaneously measure, and separate, the SZ effect and point source
contamination. (Some of these advantages are also available to large
focal-plane arrays, which form another promising avenue for future SZ
instruments; see eg Golwala, this volume).

To observe the SZ effect efficiently, however, the interferometer must
have baselines which are well matched to the angular scale of the
cluster. Each pair of antennas in an interferometer responds only to a
narrow range of fourier components of the sky brightness, centred at
$\lambda/D$ where $D$ is the antenna separation, and of width $\sim
\lambda/d$ where $d$ is the antenna size. (More exactly, the shape of
the response function in the fourier plane is the autocorrelation of
the illumination function of the antennas.) Most clusters at moderate
to high redshift have angular sizes of a few arcminutes, and hence
most of their power in the fourier plane is on scales of a few hundred
to a thousand wavelengths (see Figure 1). The
interferometers that have provided most of the SZ detections to date,
the Ryle Telescope in Cambridge and the OVRO and BIMA arrays in
California, have minimum baselines of at least 500 wavelengths
(limited by the size of the antennas), and thus resolve out most of
the flux density of the cluster. For efficient SZ surveys, telescopes
with smaller antennas and hence shorter minimum baselines are
required. This also has the beneficial effect of increasing the
instantaneous field of view of the telescope, further enhancing the
survey speed.

\begin{figure}[t!]
\plotone{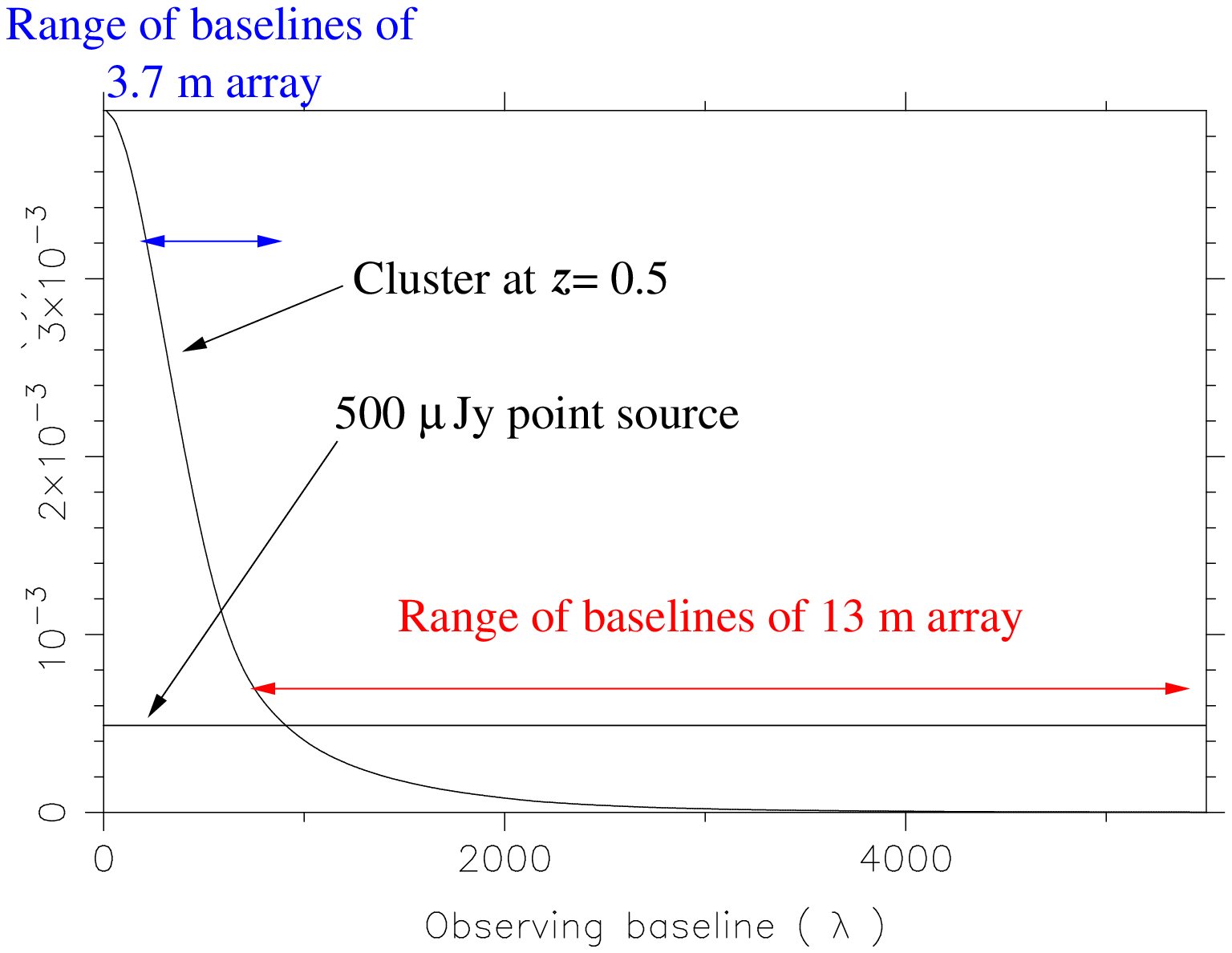}
\caption{The 15 GHz flux density due to a massive cluster at $z \sim
0.5$ as a function of interferometer baseline. Also shown are the
ranges of baselines of the two arrays of AMI, and the flux density of
a typical confusing radio source.}
\label{grainge-plot}
\end{figure}


\begin{figure}
\plotone{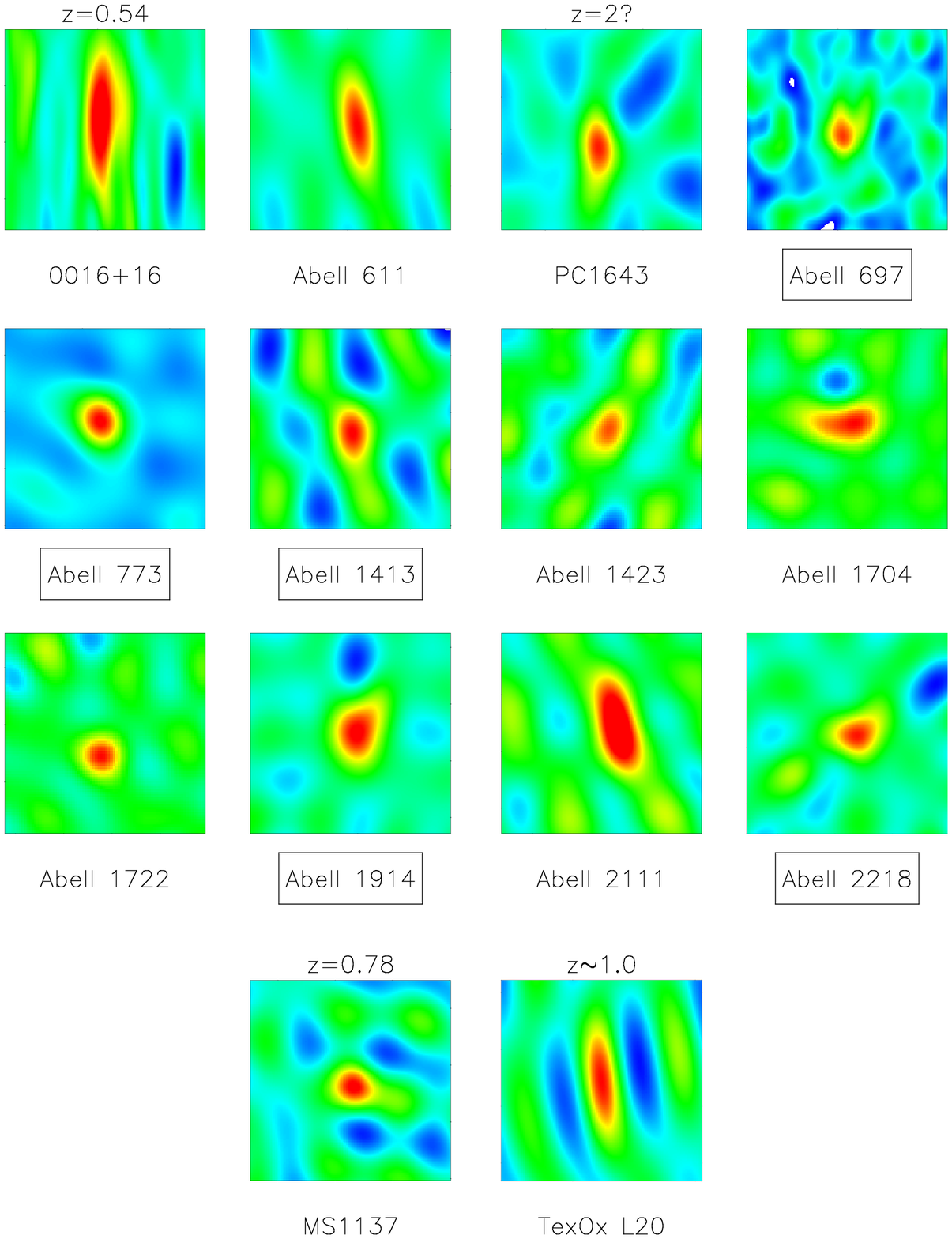}
\caption{A selection of SZ images made with the Ryle Telescope, of
clusters ranging from $z=0.14$ to $z\sim 1$. Because of the east-west
arrangement of the RT antennas, the point spread function is elongated
in the declination direction by cosec(declination). The clusters with
outlined names are those used for a determination of $H_0$ from a
complete X-ray selected sample by Jones et al (2001).}
\label{szpics}
\end{figure}

\section{SZ observations with the Ryle Telescope}

The Ryle Telescope (RT) is an east-west synthesis interferometer with
eight 13-m antennas, of which four are fixed at 1.2~km spacings and
four are mobile on a railway track. Originally constructed to make
high resolution images of radio galaxies (Ryle 1972), it has been
adapted for high brightness sensitivity observations at 2~cm
wavelength using the five antennas that can be close packed, and made
the first interferometric detection of the SZ effect (Jones et al
1993). Figure 2 shows some of the SZ detections that have
been made subsequently. We have recently used five of these, in
conjunction with X-ray data, to produce an estimate of the Hubble
constant in which a major systematic uncertainty of the SZ/X-ray
distance method, orientation bias, can be carefully addressed (Jones
et al 2001).

All but two of these SZ detections have been of objects selected
initially by optical or X-ray surveys, the completeness of which are
limited, using current techniques, to relatively low redshift, $z \la
0.3$. The other two were attempts to find high-redshift clusters by
means of `signposts' of clustered high-redshift objects. PC1643+4631
is a pair of QSOs at $z=3.8$ with an apparent associated CMB decrement
(Jones et al 1997). However, this observation has not been confirmed
by independent measurements (Holzapfel et al 2000) and must be
considered tentative. TOC J0233+3021, however, is a confirmed cluster
of galaxies at $z \sim 1$. The signpost in this case was a cluster of
faint radio sources in the NVSS catalogue with no optical counterparts
in POSS; an RT observation revealed a significant CMB decrement
(labelled TexOx~L20 in Figure 2), and subsequent deep optical and near
infra-red imaging shows clustered galaxies with colours consistent
with the redshift estimate (Cotter et al 2001). The SZ data alone show
the cluster mass to be $\sim 5 \times 10^{14} \rm M_{\odot}$.

Such observations are useful in establishing the existence of massive
high-redshift systems; however, to do more quantitative science with
clusters requires a proper survey. For this the RT is simply not
adequate. It has a bandwidth of 350 MHz and system temperature of 60 K
(considered impressive when it was designed in 1985!), giving a
sensitivity of 72 mJy s$^{-1/2}$ on 4 arcmin scales in a 6 arcmin
field. With these parameters, a useful survey---say 1 square degree
detecting clusters to $y=10^{-5}$---would take the RT $\sim 80$ years.

\section{Design parameters of AMI}

The design requirements for an SZ survey instrument are: maximum
sensitivity, hence low system temperature and high bandwidth; shortest
baselines of $\sim 200\lambda$ (confusion from primary CMB
anisotropies becomes a problem at shorter baselines); and ability to
cope with foregrounds. The first question is the choice of observing
frequency---this controls all subsequent aspects of the design.

\begin{figure}[t!]
\plotone{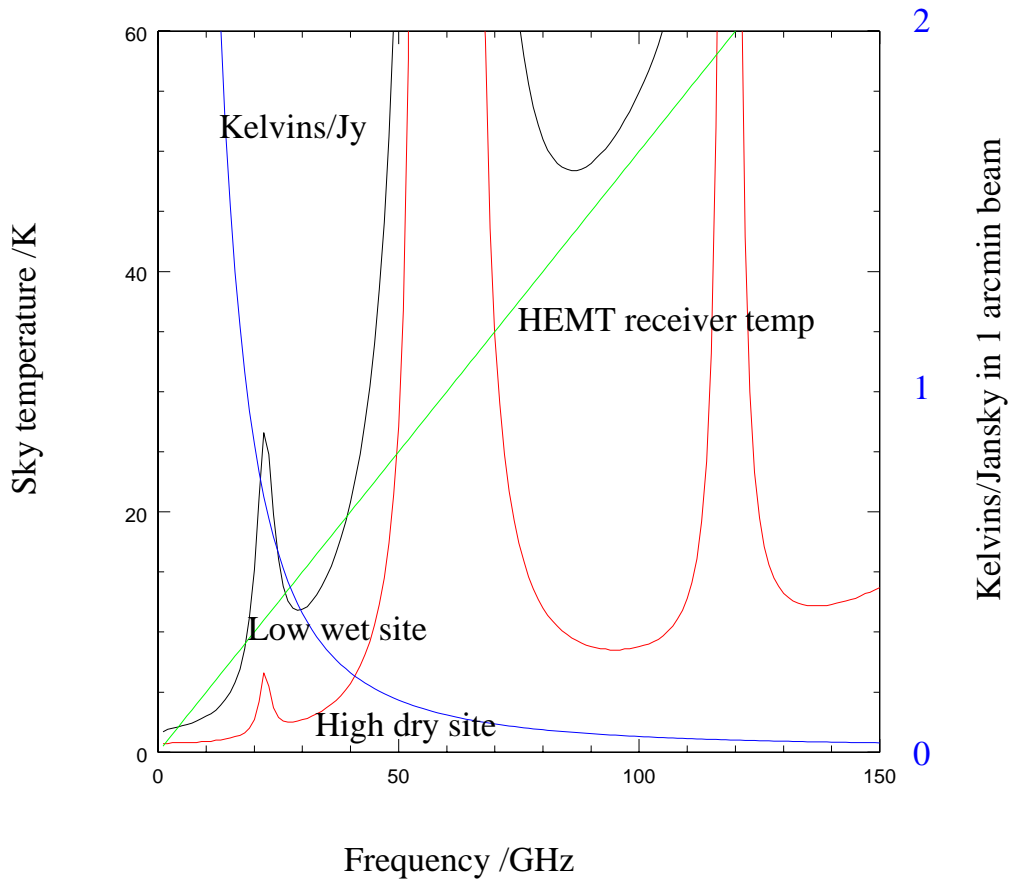}
\caption{Contributions to system temperature as a function of
frequency; the atmospheric emission temperature from a sea level site
such as Cambridge and a high dry site such as Mauna Kea are shown,
along with the typical noise temperature of a HEMT amplifier; the
total system noise is the sum of the amplifier and sky noise (plus
other contributions from losses, spillover, the CMB etc). The other
line (and scale on right) shows the flux/temperature conversion for an
unresolved source in a 1 arcmin beam.}
\end{figure}

Figure 3 shows the contributions to the system temperature of
a telescope due to the atmosphere, at wet and dry sites, and from
typical HEMT recievers. Adding these two contributions together, it
can be seen that there is a clear advantage of operating at low
frequency, even from a relatively poor site. However, this is offset
by the other feature shown on the plot, the temperature contribution
of a source of given flux density, which falls as approximately
$\nu^{-2}$. Since extragalactic point sources are the main contaminant
for SZ observations, this is an important consideration. Several
strategies are possible. One method, adopted by AMiBA, (see Lo, this
volume), is to operate at high frequency, $\nu \sim 100$~GHz, where
source contamination is minimised, and to offset the increased system
temperature by going to a high, dry site and maximising the number of
antennas and the bandwidth. An alternative is to exploit the low
system temperature available at low frequency, but to provide
sufficient flux sensitivity at higher resolution to detect and then
remove the point sources. This is the approach we adopt with AMI,
since we have the large antennas of the Ryle Telescope available to do
source subtraction. Operating at 15~GHz from Cambridge it is possible
to achieve total system temperatures below 25~K, with the cost and
logistical advantages of working only a few km from our home
institution. A similar approach is adopted by the SZA (Holder et al
2000), using a new array of small antennas in conjunction with the
existing OVRO interferometer, at 30~GHz.

With the wavelength fixed at $\lambda \simeq 2$~cm, the remaining
design parameters are fairly constrained. The available bandwith is
the $K_{\rm a}$ waveguide band of 12--18~GHz; the minimum required
baseline of $\sim 200 \lambda$ fixes the antenna size at $\sim
4$~m. We use 3.7-m diameter antennas, for which single-piece spun
reflectors are available at low cost.  The number of these antennas is
fixed by the flux sensitivity available for source subtraction from
the Ryle Telescope and the 15~GHz source counts (Taylor et al 2001) to
about ten, ie a collecting area equal to about one RT antenna. In
order to maximize the collecting area available for source
subtraction, we plan to move the three distant fixed RT antennas to
new fixed positions close to the compact array. This also has the
advantage of creating a 2-dimensional array, with much improved beam
shape at low declinations.


\subsection{The correlator}

To maximize sensitivity it is important to use as large a bandwidth as
possible. Previous high-bandwidth continuum correlators have used IF
bandwidths of $\sim ~1$~GHz, with either a single channel (eg the
VSA), or covering the entire RF band by splitting the RF band into
many IF channels (eg the CBI/DASI correlator, Padin et al). In
order to maximize the AMI bandwidth while minimizing the cost of both
the correlator and the IF system, we have been developing a correlator
to cover the entire available 6~GHz bandwidth in a single IF
channel. This design combines the recent cheap availability of the
required active components (amplifiers, switches, detectors,
voltage-variable attenuators) at frequencies above 10~GHz with the
relative ease of design of passive components (splitters, phase-shift
networks) at low {\em fractional} bandwidth. The correlator thus uses
an IF band of 6--12~GHz, with frequency resolution provided by an
8-channel multiple-lag design. An entire multi-channel correlator for
one baseline is integrated on to a single substrate. The correlation
scheme used is the phase-switched add-and-square type introduced by
Ryle (1952), also used in both CAT (Robson et al 1994) and the VSA,
which uses single detector diodes as the multiplying elements. For
more details, see Holler, this volume.

\begin{figure}
\plottwo{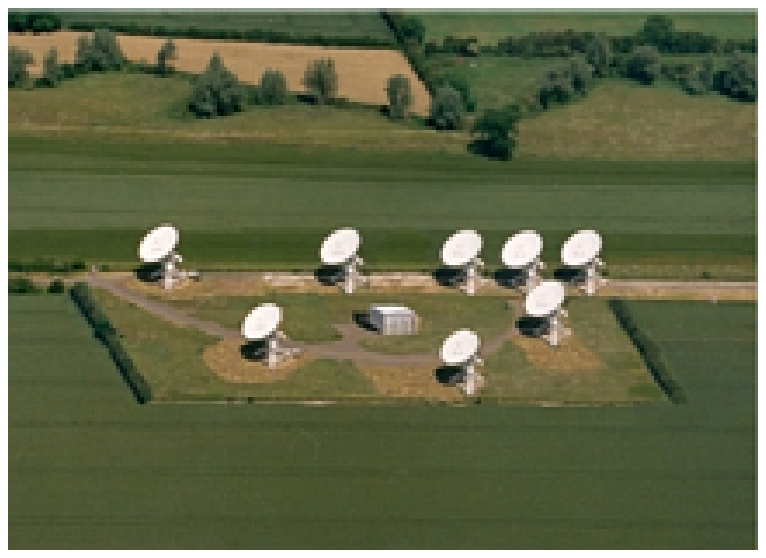}{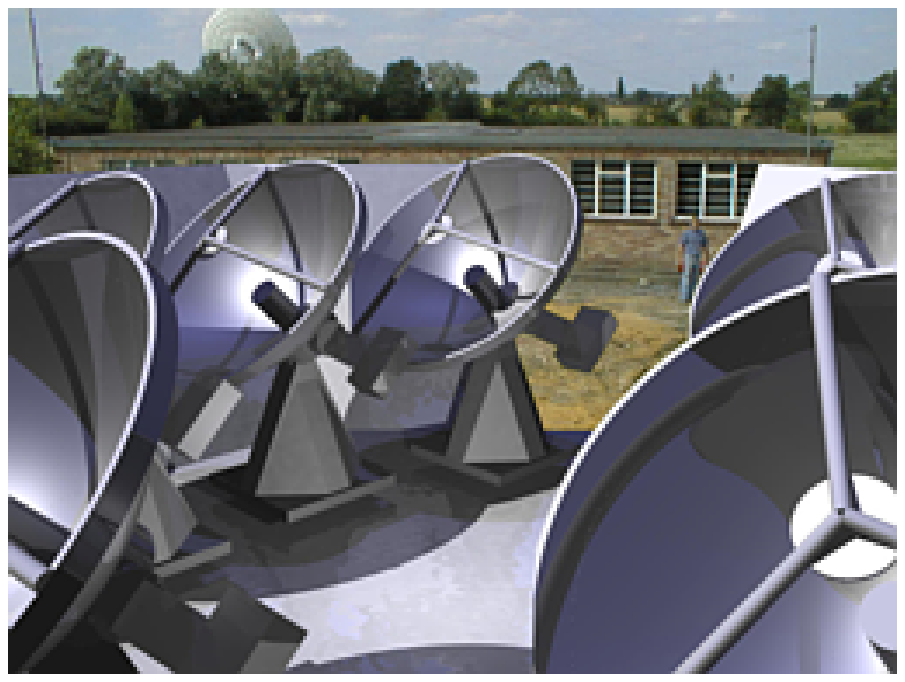}
\caption{Artists impressions of the two arrays comprising AMI.  (Left)
the 13-m antennas of the Ryle Telescope, showing the planned change to
a 2-d array. (Right) The array of 3.7-m antennas.}
\end{figure}

\section{Sensitivity}

AMI thus consists of two arrays, one of eight 13-m antennas and one of
ten 3.7-m antennas, using the same 12--18 GHz receivers and 6--12 GHz
correlator. The small array provides most of the sensitivity to the SZ
effect and the large array most of the sensitivity to point sources,
although there is some overlap. The flux sensitivities are 2 mJy
s$^{-1/2}$ and 20 mJy s$^{-1/2}$ for the large and small arrays, in 6
and 20 arcmin fields of view respectively. The temperature sensitivity
depends on the exact array configuration used; a representative number
would be about 15 $\mu$K in a week, in a 1.5 arcminute beam and 20
arcmin field of view.

More detailed calculations on the expected sensitivity to clusters and
number of clusters detected are given by Kneissl et al (2001). Figure
5 shows two simulated survey maps of about 1 square degree, under
different assumed cosmologies, showing clearly different detected
source counts.

\begin{figure}
\plottwo{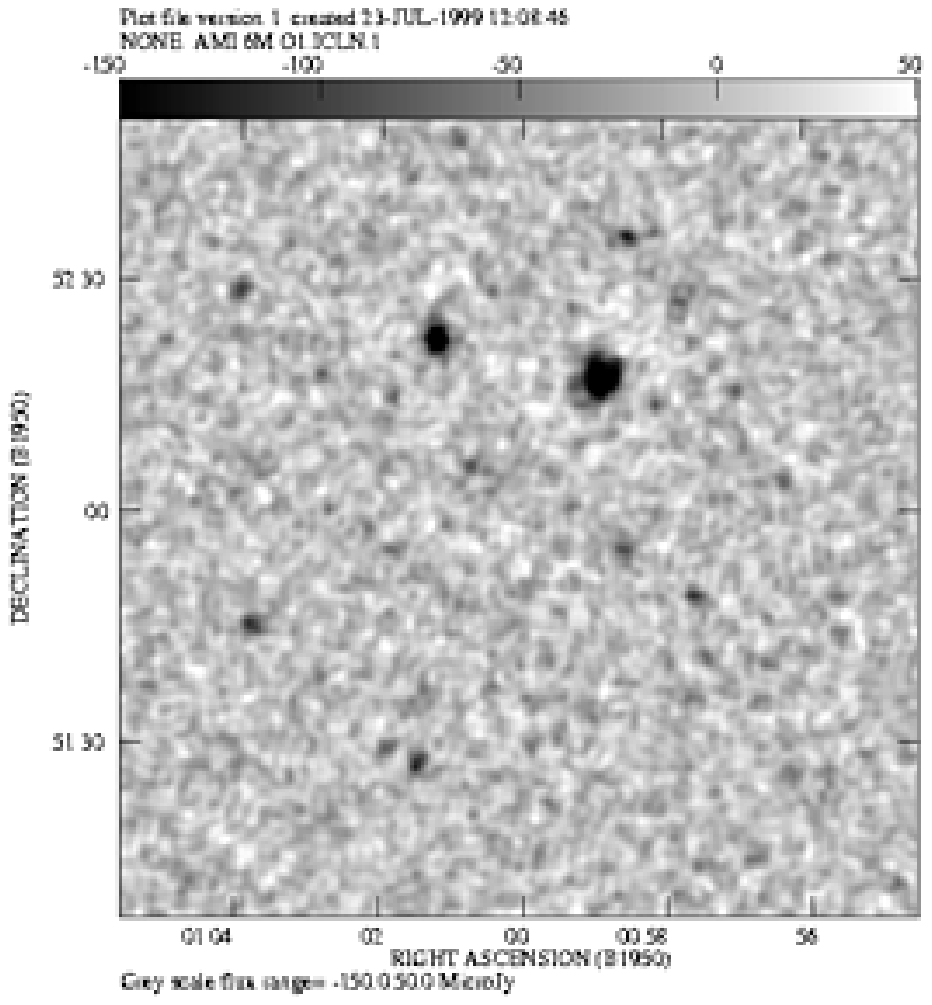}{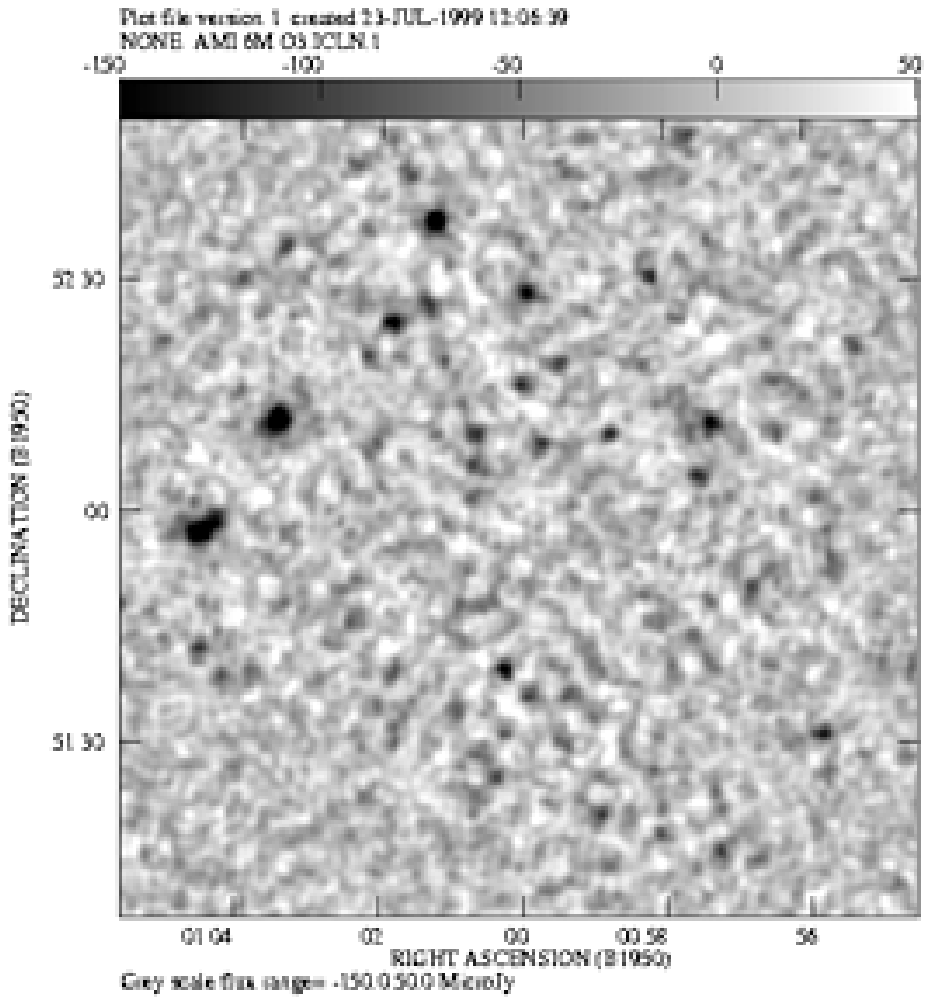}

\caption{Simulated 6-month surveys with AMI in high- (left) and low-
(right) $\Omega$ models, under pessimistic assumptions (low~$f_{\rm
b}$, low~$\sigma_8$) about the overall number of detectable
clusters. See Kneissl et al (2001) for more details.}
\end{figure}

\section{Conclusions}

The next generation of SZ instruments, including AMI as well as AMiBA
and the SZA, will provide samples of galaxy clusters with well-defined
and largely distance-independent selection criteria. These will be
invaluable in a large range of cosmological investigations. AMI itself
is now largely funded and under construction; first results are
expected by mid-2003. In recognition of the large amount of follow-up
work in other wavebands that will result from the AMI surveys, we are
proposing to make these surveys public as soon as possible after the
data are taken. Also, a significant fraction of observing time on AMI
will be open to guest observers -- details will be published when the
telescope is nearer completion.

\section{Acknowledgements}

AMI is supported by the Particle Physics and Astronomy Research
Council and the Department of Physics, University of Cambridge.

\end{document}